\documentclass[aps,twocolumn,floatfix,nofootinbib,superscriptaddress,prb]{revtex4}
\usepackage{subfigure}
\usepackage{eucal}
\usepackage{psfrag,subfigure}
\usepackage{graphicx,psfrag,subfigure}

\def\be{\begin{equation}}
\def\ee{\end{equation}}
\def\ba{\begin{eqnarray}}
\def\ea{\end{eqnarray}}

\begin{document}

\title{Seeing the magnetic monopole through the mirror of topological surface states}
\author{Xiao-Liang Qi$^1$, Rundong Li$^1$, Jiadong Zang$^2$ and Shou-Cheng Zhang}

\affiliation{Department of Physics, Stanford University,Stanford, CA
94305-4045}

\affiliation{Department of Physics, Fudan University, Shanghai,
200433, China}

\date{\today}

\begin{abstract}
Existence of the magnetic monopole is compatible with the
fundamental laws of nature, however, this illusive particle has yet
to be detected experimentally. In this work, we show that an
electric charge near the topological surface state induces an image
magnetic monopole charge due to the topological magneto-electric
effect. The magnetic field generated by the image magnetic monopole
can be experimentally measured, and the inverse square law of the
field dependence can be determined quantitatively. We propose that
this effect can be used to experimentally realize a gas of quantum
particles carrying fractional statistics, consisting of the bound
states of the electric charge and the image magnetic monopole
charge.
\end{abstract}
\pacs{}
\maketitle

The electromagnetic response of a conventional insulator is
described by a dielectric constant $\epsilon$ and a magnetic
permeability $\mu$. An electric field induces an electric
polarization, while a magnetic field induces a magnetic
polarization. Since both the electric field ${\bf E}(x)$ and the
magnetic field ${\bf B}(x)$ are well defined inside an insulator,
the linear response of a conventional insulator can be fully
described by the effective action $ S_0=\int d^3xdt (\epsilon {\bf
E}^2 - \frac{1}{\mu} {\bf B}^2)$. However, in general, another
possible term is allowed in the effective action, which is quadratic
in the electromagnetic field, contains the same number of
derivatives of the electromagnetic potential, and is rotationally
invariant; it is given by $
S_\theta=\frac{\theta}{2\pi}\frac{\alpha}{2\pi} \int d^3xdt {\bf E
\cdot B}$. Here $\alpha=e^2/\hbar c$ is the fine structure constant,
and $\theta$ can be viewed as a phenomenological parameter in the
sense of the effective Landau-Ginzburg theory. This term describes
the magneto-electric effect\cite{landau1984}, where an electric
field can induce a magnetic polarization, and a magnetic field can
induce an electric polarization.

Unlike conventional terms in the Landau-Ginzburg effective actions,
the integrand in $S_\theta$ is a total derivative term, when ${\bf
E}(x)$ and ${\bf B}(x)$ are expressed in terms of the
electromagnetic vector potential:
\begin{eqnarray}
S_\theta=\frac{\theta}{2\pi}\frac{\alpha}{16\pi} \int d^3xdt
\epsilon_{\mu\nu\rho\tau}F^{\mu\nu}F^{\rho\tau}\nonumber\\
 =\frac{\theta}{2\pi}\frac{\alpha}{4\pi}\int d^3xdt
\partial^\mu (\epsilon_{\mu\nu\rho\sigma} A^\nu\partial^\rho
A^\tau) \label{CS}
\end{eqnarray}
Furthermore, when periodic boundary condition is imposed in both the
spatial and temporal directions, the integral of such a total
derivative term is always quantized to be an integer, {\it i.e.}
$S_\theta/\hbar = \theta n$. Therefore, the partition function and
all physically measurable quantities are invariant when the $\theta$
parameter is shifted by $2\pi$ times an integer. Under time reversal
symmetry, $e^{i\theta n}$ is transformed into $e^{-i\theta n}$.
Therefore, all time reversal invariant insulators fall into two
general classes, either described by $\theta=0$ or by
$\theta=\pi$\cite{qi2008}. These two time reversal invariant classes
are disconnected, and they can only be connected continuously by
time-reversal breaking perturbations. This classification of time
reversal invariant insulators in terms of the two possible values of
the $\theta$ parameter is generally valid for insulators with
arbitrary interactions\cite{qi2008}. The effective action contains
the complete description of the electromagnetic response of
topological insulators. In Ref. \cite{qi2008}, it was shown that
such a general definition of a topological insulator reduces to the
$Z_2$ topological insulators described in Ref.
\cite{fu2006,moore2007,roy2006a} for non-interacting band
insulators, which is a 3D generalization of the quantum spin Hall
insulator in
2D\cite{kane2005A,bernevig2006a,bernevig2006d,koenig2007}. Recently,
experimental evidence of the topologically non-trivial surface
states\cite{fu2006} has been observed in $Bi_{1-x}Sb_x$
alloy\cite{hsieh2008}.

With periodic temporal and spatial boundary conditions, the
partition function is periodic in $\theta$ under the $2\pi$ shift,
and the system is invariant under the time reversal symmetry at
$\theta=0$ and $\theta=\pi$. However, with open boundary conditions,
the partition function is no longer periodic in $\theta$, and time
reversal symmetry is generally broken, but only on the boundary,
even when $\theta=(2n+1)\pi$. Ref. \cite{qi2008} gives the following
physical interpretation. Time reversal invariant topological
insulators have a bulk energy gap, but have gapless excitations with
an odd number of Dirac cones on the surface. When the surface is
coated with a thin magnetic film, time reversal symmetry is broken,
and an energy gap also opens up at the surface. In this case, the
low energy theory is completely determined by the surface term in
Eq. (\ref{CS}). Since the surface term is a Chern-Simons term, it
describes the quantum Hall effect on the surface. From the general
Chern-Simons-Landau-Ginzburg theory of the quantum Hall
effect\cite{zhang1989}, we know that the coefficient
$\theta=(2n+1)\pi$ gives a quantized Hall conductance of
$\sigma_{xy}=(n+\frac{1}{2})e^2/h$. This quantized Hall effect on
the surface is the physical origin behind the topological
magneto-electric (TME) effect. Under an applied electric field, a
quantized Hall current is induced on the surface, which in turn
generates a magnetic polarization, and vice versa.

In this work, we point out a profound manifestation of the TME
effect. When a charged particle is brought close to the surface of a
topological insulator, a magnetic monopole charge is induced as a
mirror image of the electric charge. The full set of electromagnetic
field equations can be obtained from the functional variation of the
action $S_0+S_\theta$, and they are similar to the equations of
axion electro-dynamics in particle physics\cite{wilczek1987}. They
can be presented as conventional Maxwell's equations, but with the
modified constituent equations describing the TME
effect\cite{qi2008}:
\begin{eqnarray}
{\bf D}={\bf E}+4\pi{\bf P}-2\alpha{P_3}{\bf B}\ \ ,\ \  {\bf
H}={\bf B}-4\pi{\bf M}+2\alpha{P_3}{\bf E}
\end{eqnarray}
where $P_3(x)=\theta(x)/2\pi$ is the electro-magnetic polarization
introduced in Ref. \cite{qi2008}. It takes the value of $P_3=0$ in
vacuum or conventional insulators, and $P_3=\pm 1/2$ in topological
insulators, with the sign determined by the direction of the surface
magnetization.

Now consider the geometry as shown in Fig.1. The lower half space
$(z<0)$ is occupied by a topological insulator with a dielectric
constant $\epsilon_2$ and a magnetic permeability $\mu_2$, while the
upper half space $(z>0)$ is occupied by a conventional insulator
with a dielectric constant $\epsilon_1$ and a magnetic permeability
$\mu_1$. A point electric charge $q$ is located at $(0,0,d)$ with
$d>0$. The Maxwell's equation, along with the modified constituent
equations and the standard boundary conditions, constitute a
complete boundary value problem. To solve this problem, the method
of images \cite{jackson1999} can be employed. Let us assume in the
lower half space the electric field is given by an effective point
charge $q/\epsilon_1$ and an image charge $q_1$ at $(0,0,d)$,
whereas the magnetic field is given by an image magnetic monopole
$g_1$ at $(0,0,d)$. In the upper half space the electric field is
given by an electric charge $q/\epsilon_1$ at $(0,0,d)$ and an image
charge $q_2$ at $(0,0,-d)$, the magnetic field is given by an image
magnetic monopole $g_2$ at $(0,0,-d)$. It is easily seen that the
above ansatz satisfies the Maxwell's equation on each side of the
boundary. At the boundary $z=0$ the solution is then matched
according to the standard boundary condition, giving
\begin{eqnarray}
q_1&=&q_2 =
\frac{1}{\epsilon_1}\frac{(\epsilon_1-\epsilon_2)(1/\mu_1+1/\mu_2)-4\alpha^2P_3^2}
{(\epsilon_1+\epsilon_2)(1/\mu_1+1/\mu_2)+4\alpha^2P_3^2}q\nonumber \\
g_1&=&-g_2 = -\frac{4{\alpha}P_3}
{(\epsilon_1+\epsilon_2)(1/\mu_1+1/\mu_2)+4\alpha^2P_3^2}q.\label{imagemonopole}
\end{eqnarray}
To see the physics more clearly, we will first take
$\epsilon_1=\epsilon_2=\mu_1=\mu_2=1$ below, and then recover the
$\epsilon_{1,2},~\mu_{1,2}$ when discussing the experimental
proposals later. The solution above shows that for an electric
charge near the surface of a topological insulator, both an image
magnetic monopole and an image electric charge will be induced, as
compared with conventional electromagnetic media where only an
electric image charge will be induced. It is remarkable that the
magnitude of the image magnetic monopole and image electric charge
satisfy the relation $q_{1,2}=\pm(\alpha{P_3})g_{1,2}$. This is just
the relation $q=(\theta/2\pi)g$ derived by Witten\cite{witten1979},
for the electric and magnetic charges of a dyon inside the $\theta$
vacuum, with $\theta/2\pi=\pm{P_3}$ here.

The physical origin of the image magnetic monopole is understood by
rewriting part of the Maxwell's equation as
\begin{equation}
\nabla\times{\bf B}=2\alpha{P_3}\delta(z)\hat{\bf n}\times{\bf E}.
\end{equation}
with $P_3=\pm 1/2$ the value for the topological insulator. The
right hand side of the above equation corresponds to a surface
current density ${\bf j}=\sigma_{xy}(\hat{\bf n}\times{\bf E})$,
which is induced by the in-plane component of the electric field and
is perpendicular to it. This current is nothing but the quantized
Hall current mentioned earlier. For the problem under consideration,
the surface current density is calculated as
\begin{equation}
{\bf
j}=P_3\left(\frac{e^2}{h}\right)\left(\frac{q}{1+\alpha^2{P_3}^2}\right)\frac{r}{(r^2+d^2)^{\frac{3}{2}}}\hat{\bf
e}_\varphi,
\end{equation}
which is circulating around the origin as shown in the inset of
Fig.1. Physically, this surface current is the source that induces
the magnetic field. On each side of the surface, the magnetic field
induced by the surface current can be viewed as the field induced by
an image magnetic monopole on the opposite side.

\begin{figure}
        \label{fig1}
          \includegraphics[width=0.44\textwidth]{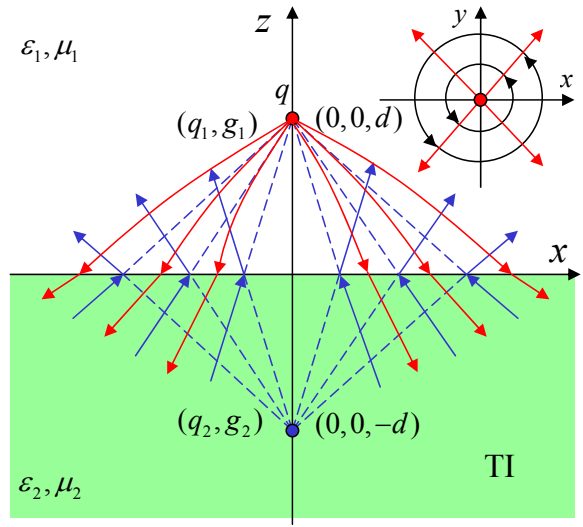}
\caption{Illustration of the image electric charge and the image
monopole of a point-like electric charge. The lower half space is
occupied by a topological insulator (TI) with dielectric constant
$\epsilon_2$ and magnetic permeability $\mu_2$. The upper half space
is occupied by a topologically trivial insulator (or vacuum) with
dielectric constant $\epsilon_1$ and magnetic permeability $\mu_1$.
A point electric charge $q$ is located at $(0,0,d)$. Seen from the
lower half space, the image electric charge $q_1$ and magnetic
monopole $g_1$ are at $(0,0,d)$. Seen from the upper half space, the
image electric charge $q_2$ and magnetic monopole $g_2$ are at
$(0,0,-d)$. The red (blue) solid lines represent the electric
(magnetic) field lines. The inset is a top-down view showing the
in-plane component of the electric field at the surface (red arrows)
and the circulating surface current (black circles).}
  \end{figure}

From the above calculation we clearly see that the image magnetic
monopole field indeed has the correct magnetic field dependence
expected from a monopole, and it can be controlled completely
through the position of the electric charge. Since we started with
the Maxwell's equation, which includes ${\bf \nabla \cdot B}=0$, the
magnetic flux integrated over a closed surface must vanish. We can
indeed check that this is the case by considering a closed surface,
for example a sphere with radius $a$, which encloses a topological
insulator. Inside the closed surface, there is not only a image
magnetic monopole charge, but also a line of magnetic charge density
whose integral exactly cancels the point image magnetic monopole.
However, when the separation between the electric charge and the
surface, $d$, is much smaller than the spherical radius $a$, the
magnetic field is completely dominated by the image magnetic
monopole, and the contribution due to the line of magnetic charge
density is vanishingly small. Therefore, we propose here to
experimentally observe the magnetic monopole in the same sense as we
can experimentally observe other fractionalization, or
de-confinement, phenomena in condensed matter physics. In any closed
electronic system, the total charge must be quantized to be an
integer. However, one can separate fractionally charged elementary
excitations arbitrarily far from each other, so that fractional
charge can have well defined meaning locally. Similar situation
occurs in spin-charge separation. While the total charge and the
total spin of a closed system must be linked to each other, spin and
charge can occur as well separated local excitations. In our case,
as long as $d$ is much smaller than the radius of curvature of a
topological surface $a$, the local magnetic field is completely
determined by a single image magnetic monopole.

\begin{figure}
        \label{fig2}
          \includegraphics[width=0.44\textwidth]{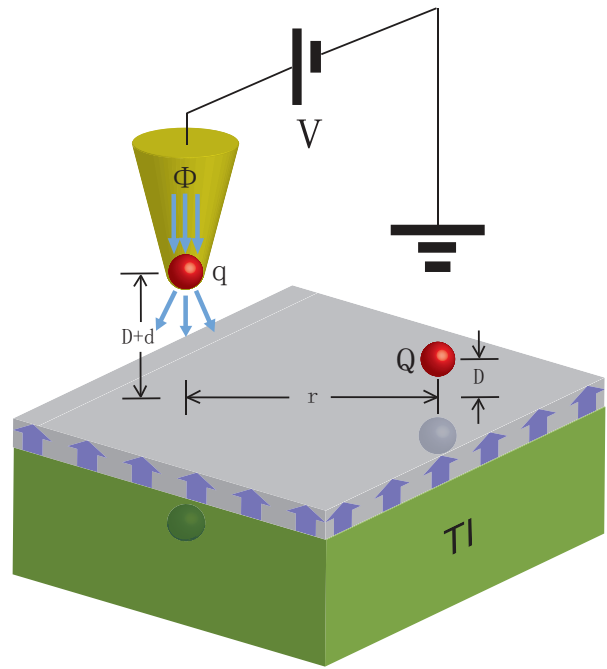}
\caption{Illustration of the experimental setting to measure the
image monopole. A magnetic layer is deposited on the surface of the
topological insulator, as indicated by the layer with blue arrows.
(The same layer is drawn in Fig. 3 and 4.) A scanning MFM tip
carries a magnetic flux $\phi$ and a charge $q$. A charged impurity
is confined on the surface with charge $Q$ and distance $D$ to the
surface. By scanning over the voltage $V$ and the distance $r$ to
the impurity, the effect of the image monopole magnetic field can be
measured (see text).}
  \end{figure}

Such an image monopole can be observed experimentally by a Magnetic
Force Microscope (MFM). Consider the surface of the topological
insulator with a localized charged impurity, as shown in Fig. 2. A
scanning MFM tip can be applied to detect the magnetic field
distribution of the image monopole. However, the charge of the
impurity also generates an electric force to the tip. Here we show
that the contribution of the image monopole can be distinguished
from other trivial contributions by scanning both the tip position
$r$ and the tip voltage $V$. For a given position $r$, we define
$f_{\rm min}(r)$ as the minimal force applied to the tip when
scanning the voltage. Denote the distance of the charged impurity to
the surface as $D$, in the limit of $r\gg D$ the conventional charge
interaction leads to a $1/r^6$ dependence of $f_{\rm min}(r)$. On
comparison, the image monopole magnetic field leads to more dominant
contribution
\begin{eqnarray}
f_{\rm min}(r)\simeq \frac{4\alpha
P_3}{\left(1+\epsilon_2/\epsilon_1\right)\left(1/\mu_1+1/\mu_2\right)-4\alpha^2P_3^2}\frac{Q\phi}{r^3}
\end{eqnarray}
in which $Q$ is the impurity charge and $\phi$ is the net flux of
the magnetic tip. For the estimated parameters $\epsilon_2\simeq
100$ for ${\rm Bi_{1-x}Sb_x}$ alloy\cite{landolt1998},
$\epsilon_1=1,\mu_1\simeq \mu_2\simeq 1,~\phi\simeq 2.5hc/e$ and a
typical distance $r=50{\rm nm}$, the force is $f_{\rm min}(r)\simeq
0.3pN/\mu m$, which is observable in the present experiments.

\begin{figure}
        \label{fig3}
          \includegraphics[width=0.44\textwidth]{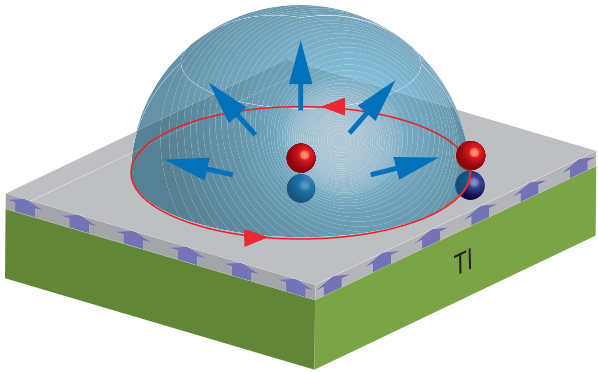}
\caption{Illustration of the fractional statistics induced by the
image monopole effect. Each electron forms a ``dyon" with its image
monopole. When two electrons are exchanged, a Aharonov-Bohm phase
factor is obtained, which is determined by half of the image
monopole flux independent of the exchange path, leading to the
phenomenon of statistical transmutation.}
  \end{figure}

When we consider more than one point charges on the surface, the
existence of such an image monopole has another important
consequence. When an electron is moving on the surface of the
topological insulator, it is always followed by its image monopole.
In other words, each electron behaves like a bound state of charge
and monopole, which is known in the high energy physics as a
``dyon"\cite{witten1979}. When two dyons are exchanged, each of them
will obtain an Aharonov-Bohm (AB) phase due to the magnetic field of
the other one. Interestingly, the net AB phase obtained by the
two-particle system during an exchange process is independent of the
path of the particles on the 2D plane, which thus can be interpreted
as a statistical angle of the dyon. Therefore, this setup provides a
condensed matter realization of the anyon
concept\cite{leinaas1977,wilczek1982} in the absence of any external
magnetic field. This effect also provides a mechanism of statistical
transmutation in $(3+1)$ dimensions\cite{jackiw1976,tHooft1976}. If
each dyon has charge $q$ and monopole flux $g$, then the statistical
angle is given by
\begin{eqnarray}
\theta=\frac{gq}{2\hbar c}
\end{eqnarray}
For example, the binding of an electron with charge $e$ and a
monopole with flux $hc/e$ leads to $\theta=\pi$, in which case the
dyon is a boson. However, there is an interesting difference between
the dyon we studied here and that in high energy physics. In high
energy physics the charged particle and monopole are point like
particles moving in the 3D. Consequently, the monopole flux must be
quantized in unit of $hc/e$, and correspondingly, the statistical
angle $\theta$ can only be $0$ or $\pi$ modular $2\pi$. This is
consistent with the fact that there is no anyonic statistics in
$(3+1)$ dimensions. On comparison, our dyon can only be defined if
an electron is moving closed to the surface of a topological
insulator. Consequently, the ``dyon" is always confined on a 2D
surface. In this case, the flux of the image monopole does not have
to be quantized, and correspondingly, the ``dyon" can have anyonic
statistics. According to Eq. (\ref{imagemonopole}), the statistical
angle of an electron-induced dyon is
\begin{eqnarray}
\theta=\frac{2{\alpha}^2P_3}
{(\epsilon_1+\epsilon_2)(1/\mu_1+1/\mu_2)+4\alpha^2P_3^2}\label{statangle}
\end{eqnarray}
For $P_3=1/2$ and $\epsilon_1,\mu_1,\mu_2\sim 1,~\epsilon_2\sim
100$, we obtain $\theta\simeq \alpha^2/200\simeq 2.6\times
10^{-7}{\rm rad}$. Though the statistical angle is quite small, it
is physically observable. Consider the geometry as shown in Fig. 4.
A quasi 1D superconducting ring and a metallic island surrounded by
the ring are deposited on top of the topological insulator surface
(which has already been gapped by a magnetic layer). By tuning the
gate voltage of the central island, the number of electrons $N$ on
the island can be tuned. Due to the statistical angle $\theta$ given
by Eq. (\ref{statangle}), each electron in the central island
induces a flux of $\theta$ seen by the electrons in the ring.
Consequently, the net flux through the ring is $N\theta$ in unit of
the flux quanta, which generates a supercurrent in the
superconducting ring as in standard Superconducting Quantum
Interference Device (SQUID). For a typical electron density of
$n=10^{11}/{\rm cm}^{2}$ and island size $R=1{\rm \mu m}$, the net
magnetic flux is $n\theta\pi R^2{\hbar c}/e\simeq 2.6\times
10^{-4}hc/2e$ and the corresponding magnetic field is
$B=n\theta\hbar c/e\simeq 1.7\times 10^{-3}{\rm G}$, which is
observable for SQUID today. It is possible to have some other
topological insulator material with a smaller dielectric constant
$\epsilon_2$, in which the statistical angle of dyon can be larger.

In summary, in this paper we have demonstrated that the topological
surface states of a 3D topological insulator act as a mirror which
images an electron as a monopole. Such a transmutation between
electric field and magnetic field is a direct manifestation of the
TME effect discussed in Ref. \cite{qi2008}. Due to this effect, a 2D
electron gas in the neighborhood of the surface will becomes a
``dyon" gas with fractional statistics. Exotic particles such as the
magnetic monopole, dyon, anyon, and the axion have played
fundamental roles in our theoretical understanding of quantum
physics. Experimental observation of these exotic particles in table
top condensed matter systems could finally reveal their deep
mysteries.

\begin{figure}
        \label{fig4}
          \includegraphics[width=0.44\textwidth]{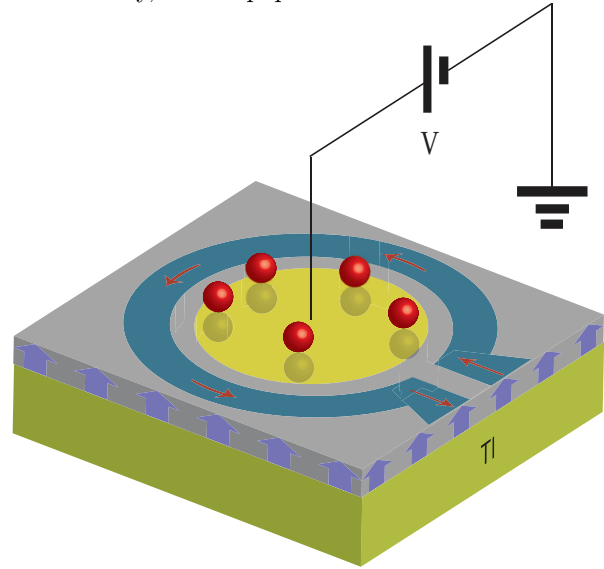}
\caption{Illustration of the experimental proposal measuring the
fractional statistics of the dyons. When a gate voltage is applied
to the central metallic island, the number of electrons in the
central island can be tuned, which in turn changes the net flux
threaded in the superconducting ring and leads to a supercurrent.}
  \end{figure}

 We wish to thank T. L. Hughes, L. Luan and O. M. Auslaender for
insightful discussions. This work is supported by the NSF through
the grants DMR-0342832, and by the US Department of Energy, Office
of Basic Energy Sciences under contract DE-AC03-76SF00515, and by
the Ministry of Education of China under the Grant No. B06011.
\bibliography{Monopole}

\end{document}